\documentclass[aps,prl,citeautoscript,twocolumn,reprint,longbibliography,floatfix,superscriptaddress]{revtex4-2}
\usepackage{graphicx, amssymb, amsmath, epsf, bm,cprotect, comment, physics}
\epsfclipon
\usepackage[utf8]{inputenc}
\usepackage{mathrsfs}
\usepackage{mathtools}
\usepackage{color}
\usepackage{url}
\usepackage{bm}
\usepackage{ulem}
\usepackage{layout}

\newcommand{\expct}[1]{\langle{#1}\rangle}

\renewcommand{\eqref}[1]{Eq.\,(\ref{#1})}

\newcommand{\figref}[1]{Fig.\,\ref{#1}}

\newcommand{\supfigref}[1]{Fig.\,\ref{#1}}

\newcommand{\red}[1]{#1}

\usepackage{xr} 
\makeatletter
\newcommand*{\addFileDependency}[1]{
  \typeout{(#1)}
  \@addtofilelist{#1}
  \IfFileExists{#1}{}{\typeout{No file #1.}}
}
\makeatother

\begin{document}

\title{Enzyme as Maxwell's Demon: Steady-state Deviation from Chemical Equilibrium by Enhanced Enzyme Diffusion}

\author{Shunsuke Ichii}
\email{ichii@ubi.s.u-tokyo.ac.jp}
\affiliation{Department of Physics,\! The University of Tokyo,\! 7-3-1 Hongo,\! Bunkyo-ku,\! Tokyo 113-0033,\! Japan}%
\affiliation{Center for Biosystem Dynamics Research, RIKEN, 6-2-3 Furuedai, Suita, Osaka 565-0874, Japan}%

\author{Tetsuhiro S. Hatakeyama}
\email{hatakeyama@elsi.jp}
\affiliation{Earth-Life Science Institute, Institute of Science Tokyo, 2-12-1 Ookayama, Meguro-ku, Tokyo 152-8550, Japan}%

\author{Kunihiko Kaneko}
\email{kunihiko.kaneko@nbi.ku.dk}
\affiliation{Earth-Life Science Institute, Institute of Science Tokyo, 2-12-1 Ookayama, Meguro-ku, Tokyo 152-8550, Japan}%
\affiliation{Niels Bohr Institute, University of Copenhagen, Blegdamsvej 17, 2100 Copenhagen, Denmark}%

\date{\today}

\begin{abstract}
Enhanced enzyme diffusion (EED), in which the diffusion coefficient of an enzyme transiently increases during catalysis, has been extensively reported experimentally. We numerically and analytically demonstrate that such enzymes can act as Maxwell's demons. They use their enhanced diffusion as a memory of the previous catalytic reaction, to gain information and drive steady-state chemical concentrations away from chemical equilibrium. Our theoretical analysis identifies the conditions for this process, highlighting the functional role of EED and its relevance to cellular systems.
\end{abstract}

\maketitle

While enzymes were traditionally considered only in terms of catalytic activities, recent studies have unveiled novel charactaristics, termed as enhanced enzyme diffusion (EED), in which the diffusion coefficient of an enzyme increases when it catalyzes a reaction \cite{Feng2020-ry}. 
This phenomenon has been reported in various enzymes, from urease \cite{Muddana2010-yi, Sengupta2013-hq, Dey2015-rl, Jee2018-ys, Jee2018-hp, Jee2019-ex, Lin2022-ct}, catalase \cite{Sengupta2013-hq, Dey2015-rl, Sun2017-hi}, glycolytic enzymes \cite{Illien2017-uo, Zhao2018-zt}, ATPase \cite{Borsch1998-ka}, DNA and RNA polymerases \cite{Yu2009-lb, Sengupta2014-gt} and so far \cite{Pavlick2013-cq, Jee2018-hp, Wang2020-np}, while some are under discussion \cite{Gunther2018-so, Zhang2018-lj, Gunther2019-zw, MacDonald2019-my, Zhang2019-bs, Chen2020-kf, Choi2022-ba}. 
Possible mechanisms behind this phenomenon are under active investigation, and various models have been proposed \cite{Golestanian2009-fp, Golestanian2010-jk, Muddana2010-yi, Sakaue2010-xc, Bai2015-ds, Illien2017-uo, Illien2017-oa, Agudo-Canalejo2018-td, Agudo-Canalejo2018-yj, Jee2019-ex, Adeleke-Larodo2019-cg, Kondrat2019-je, Feng2019-wq, Jee2020-yt, Skora2021-md, Ryabov2022-qb, Vishen2024-qu, Yancheva2025-nu}. 
While \red{its microscopic origin has not yet been settled}, the existence of EED itself is accepted, which gives a microscopic process to convert the energy of chemical reactions into the motility of enzymes \cite{Riedel2015-gz, Golestanian2015-lt, Jee2018-ys, Jee2020-yt, Krist2023-ym}. 
In this way, enzymes are microscopic thermodynamic engines. 

In general, microscopic thermodynamic engines, such as molecular motors, can cause macroscopical changes in the system. 
Indeed, EED has also been reported to move enzymes according to the gradient of substrates, which is similar to \red{phoretic motion or }chemotaxis \cite{Jee2018-ys, Zhao2018-zt, Xu2019-lf, Wang2020-np, Krist2023-ym}. 
Spatial inhomogeneity induced by EED \red{by itself} has also been discussed. 
However, since the enzymes do not change the equilibrium of chemical reactions, whether EED can change macroscopic chemical composition of reactants has not been explored. 

In this Letter, we demonstrate that EED can deviate the steady state concentration of reactants from the chemical equilibrium. 
With EED \red{by comsuming energy}, the information of a substrate catalyzed by the enzyme is stored as a change in diffusion coefficient of the enzyme, and it biases the probability of following reactions by using the stored information. 
Therefore, the steady state can deviate from the chemical equilibrium. 
In this way, enzymes can work as information thermodynamic engines like Maxwell's demon \cite{Maxwell1871-aa, Szilard1929-es, Sagawa2008-id}. 
We derived the conditions under which EED enables enzymes to act as Maxwell's demon, and found that these conditions align with experimentally observed parameters for real enzymes.

\begin{figure*}[tb]
  \centering
  \includegraphics[width=2\columnwidth]{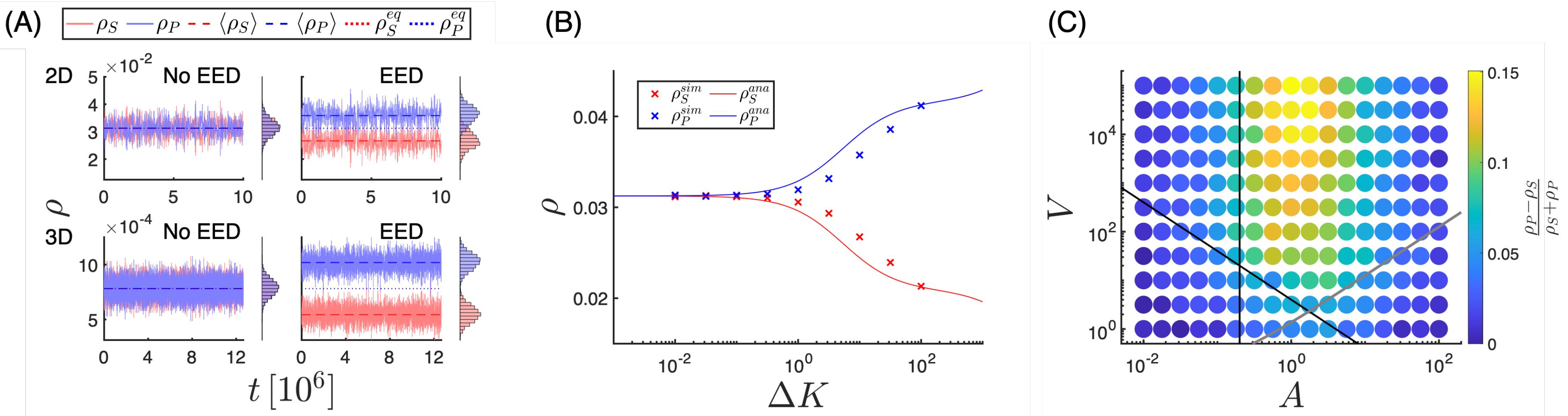}
  \caption{(A) The steady-state concentrations of $S$ and $P$, $\rho_S$ and $\rho_P$\red{, in two- and three-dimensional particle-reaction simulations}. 
  \red{The time series of $\rho_S$ (red) and $\rho_P$ (blue) are plotted for the two-dimensional case (top) and the three-dimensional case (bottom) along with histograms of the corresponding distributions.}
  The lines represent the time series of $\rho_S$ and $\rho_P$ in the simulation.
  The dotted line represents the value of $\rho_S,\rho_P$ in calculated chemical equilibrium. 
  The dashed line represents the average concentrations over time. 
  The top panel is a plot for $\Delta K = 0$ and the bottom for $\Delta K = 10$. 
  (B) $\Delta K$ dependency of the average $\rho_S$ and $\rho_P$ at steady states. 
  The cross marks represent the simulation results and the solid lines represent the analytical solution of four-state model, which is introduced later.  
  (C) Phase diagram of the deviation from equilibrium $(\rho_P-\rho_S)/(\rho_S+\rho_P)$ against the parameters $A$ and $V$. 
  The lines represent the analytically estimated boundaries where the deviation from equilibrium disappears, given by $A = \frac{2(1+\kappa)}{(\Delta K + 2) \kappa}$, $AV = 4$, and $(\Delta K A+1)(\rho_S+\rho_P) = V$. }
  \label{fig:simulationResults}
\end{figure*}

We adopt a particle-reaction model consisting of particles underlying enzymatic reactions with EED and analyze the macroscopic behavior in a steady state. 
In actual biological systems, an enormous number of reactions are interconnected. 
For the sake of simplicity, however, we focus on a single set of forward and reverse enzyme reactions $S + E \leftrightarrows P + E$, to pinpoint the macroscopic effect of EED, where $S$, $P$, and $E$ represent the substrate, product, and enzyme, respectively.

We consider a system where $N$ particles with a radius $r = 0.5$, each labeled as $S$, $E$, or $P$. 
\red{The particles are placed in a two- or three-dimensional space of size $L = 40$ with periodic boundary conditions.}
Each particle moves solely by Brownian motion. 
We represent the coordinates of particle $i$ by $\bm{x}_i$, and the time evolution is described by the overdamped Langevin equation of motion as
\begin{equation} \label{eqn:particleMove}
  \dv{t} \bm{x}_i = \sqrt{2 D_i} \bm{R}(t)
\end{equation} 
where $D_i$ is the diffusion coefficient of particle $i$, and $\bm{R}(t)$ is a white noise with $\expct{\bm{R}(t)} = 0$ and $\expct{\bm{R}(t) \bm{R}(t')} = \delta(t-t')$.

When $S$ or $P$ overlaps with $E$ particle within the radius $r$, its state changes probablistically to $P$ or $S$ with a rate of $a$ or $\kappa a$, respectively, where $\kappa$ is the ratio of the reverse reaction rate to the forward reaction rate, defined as $\kappa = \exp[\beta(\mu_S - \mu_P)]$, $\beta$ is the inverse temperature, and $\mu_S$ and $\mu_P$ are the chemical potentials of $S$ and $P$, respectively. 
As the simplest case to study the effect of EED, we assume $\mu_S - \mu_P = 0$, meaning $\kappa = 1$, unless otherwise mentioned. 
Then the concentrations of $S$ and $P$ are equal in the equilibrium state. 

For simplicity, here, internal structure of the enzyme is not considered. 
As long as the particles overlap, the reaction occurs at the same rate, regardless of their specific spatial arrangement. 

EED is implemented by a \red{temporary} increase in the diffusion coefficient of the enzyme \red{following an $S \rightarrow P$ reaction}\red{\footnote{\red{The experimentally confirmed selective increase in diffusion for exothermic reactions violates Onsager's reciprocal relations and inherently requires a non-equilibrium process.}}}. 
\red{This asymmetric enhancement of diffusion is a key feature of EED, as observed experimentally \cite{Jee2020-yt}.}
\red{Such behavior requires an external energy source, such as conjugate chemical reactions. Here, we assume its presence implicitly without specifying its mechanism, as our conclusions remain valid regardless of its exact form.}
The diffusion coefficient of particle $i$, $D_i$, is decomposed into $D_i = K_i / \nu$. 
$K_i$ is a dimensionless variable representing the motility of a particle, and $\nu$ is a viscosity parameter normalized for $K_i=1$. 
When an enzyme catalyzes the forward reaction, its $K_i$ is transiently increased by a specific value $\Delta K$, i.e., $K_i \to K_i + \Delta K$, and then decays at a rate of $\gamma$ to unity as given by 
\begin{equation} \label{eqn:motilityDecay}
  \dv{t} K_i = - \gamma (K_i - 1)
\end{equation}
\red{Thus, the memory of the enzyme’s high-motility state, acquired upon catalyzing the reaction, decays through the dissipation of heat to the environment via the solvent.}

The sum of the number of substrate and product particles is fixed at 100 and the number of enzyme particles is fixed at 500, unless otherwise mentioned. 
We denote the number density of particles as $\rho$ with its subscripts $S$, $E$, and $P$, respectively.

First, we simulated this particle-reaction model over a sufficiently long time span, to obtain the steady-state $\rho_S$ and $\rho_P$. 
As shown in \figref{fig:simulationResults}A, the ratio of $\rho_S$ to $\rho_P$ converged to the value at the chemical equilibrium, $\rho_S / \rho_P = \kappa = 1$, when $\Delta K = 0$, i.e., no EED was implemented. 
However, when $\Delta K > 0$, $\rho_P$ converged to higher concentration than $\rho_S$, indicating that the system did not converge to the chemical equilibrium. 
\red{As shown in \figref{fig:simulationResults}A, the effect of EED is more pronounced in three-dimensional simulations than in two-dimensional ones. In this Letter, however, we mainly focus on the two-dimensional case for simplicity.}
This deviation monotonically increased with $\Delta K$, as shown in \figref{fig:simulationResults}B, which indicates that EED can deviate the steady-state concentrations of $S$ and $P$ away from equilibrium. 

We examined how this deviation from equilibrium depends on the parameters of the system. 
We focus especially on the two dimensionless parameters related to time scale of the system. 
One is the ratio of the reaction rate to the decay rate of the motility, $A \coloneq a / \gamma$, which represents the relative timescale of the dissipation of motility. 
The other is the persistent length of the motility increase effect, $V \coloneq 4 \gamma \nu r^2$. 
These two parameters characterize the relation between the reaction and the EED. 
By simulating the deviation from equilibrium across the ranges of $A$ and $V$, we found that the deviation is salient only within a specific region (\figref{fig:simulationResults}C), where $A$ is close to unity and $V$ is above $10$.

\begin{figure}[tb]
  \centering
  \includegraphics[width=0.8\columnwidth]{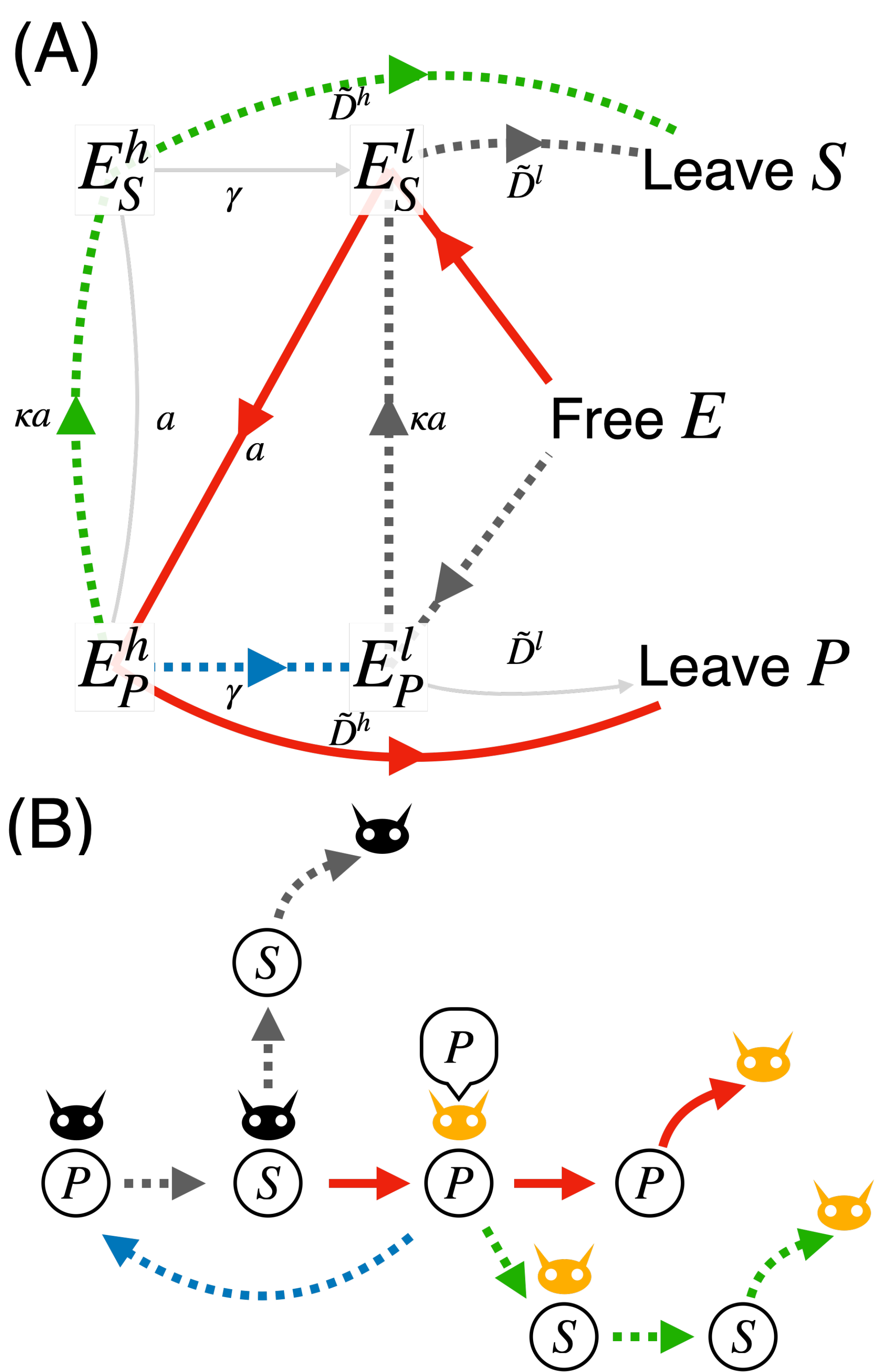}
  \caption{(A) Transition diagram of the state of $E$ with the reactants $S$ and $P$. 
  The state of the enzyme $E^i_j$ is characterized by two factors: 
  Whether $E$ overlaps with $S$ or $P$, denoted by subscripts as $i = S$ or $P$, and whether $E$ is in a high- or low-motility state, denoted by superscripts as $j = h$ or $l$. 
  (B) \red{A} schematic representation of \red{the typical pathway and its side routes where the enzyme either functions or fails to function as a Maxwell’s demon. 
  Along these pathways, the demon retains or loses the memory of the reactant in the form of its motility state, represented by the orange and black colors, respectively, before completely detaching from the reactant.} 
  \red{(A, B) The red arrows represent the typical pathway in which the enzyme successfully functions as demon, observing the state of the reactant and using that information to selectively leave the product $P$. 
  The black, blue, and green dotted arrows indicate side routes that interfere with the demon's function (see main text for details). }}
  \label{fig:demonPathways}
\end{figure}

\begin{figure}[tb]
  \centering
  \includegraphics[width=0.9\columnwidth]{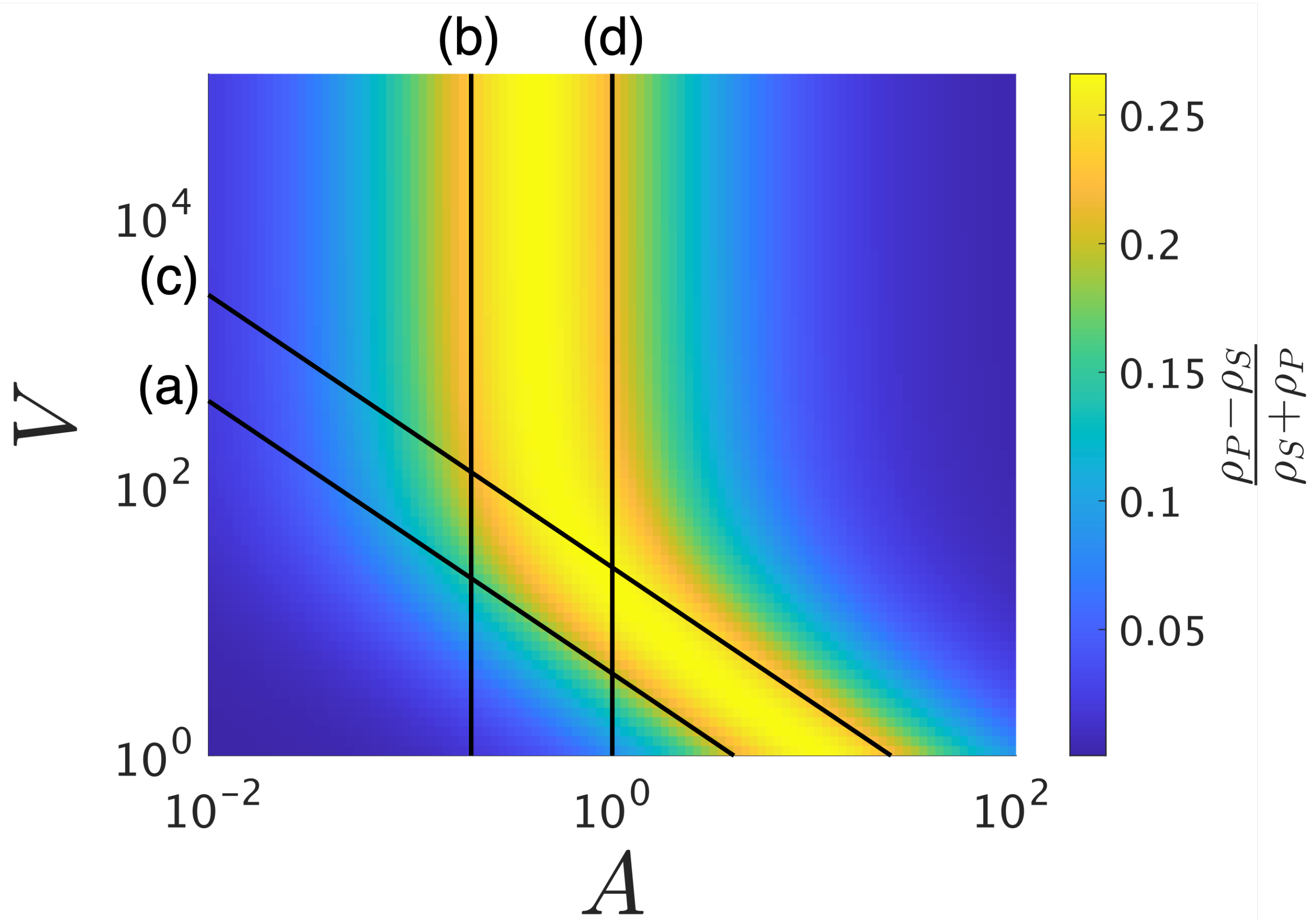}
  \caption{Phase diagram of the deviation from equilibrium to steady state for different values of $A$ and $V$ with four-state model. 
  Lines represent the boundaries where the model does not exhibit the deviation from equilibrium. 
  Line (a), given by $\tilde D^l = a$, represents the boundary with the region where the reaction is too slow, corresponding to the blue arrows in \figref{fig:demonPathways}. 
  Line (b), given by $\tilde D^h \frac{\kappa}{1+\kappa} a = \tilde D^l \gamma$, represents the boundary with the region where the high-motility state of $E$ is not sufficiently realized and has less effect, corresponding to the cyan arrows in \figref{fig:demonPathways}. 
  Line (c), given by $\kappa a = \tilde D^h$, and line (d), given by $\kappa a = \gamma$, represent the boundary with the region where the reverse reaction is too fast, corresponding to the yellow arrows in \figref{fig:demonPathways}.}
  \label{fig:analysis}
\end{figure}

\red{To understand the} mechanism \red{underlying} the deviation from the equilibrium\red{, we consider a single reactant and a
single enzyme that overlap within distance $r$, along with
the stochastic leaving process of the enzyme from the
reactant. This simplification is justified by assuming
that the enhanced diffusion decays after the enzyme leaves
the reactant and before it overlaps with another reactant again.
As long as the enzymatic reaction does not occur in cascade,
this assumption remains valid.}

\red{In the simplified model, }The state of the system is characterized by two factors: 
Whether $E$ overlaps with $S$ or $P$, and whether $E$ is in a high- or low-motility state, where, for the sake of analysis, we simplify that the diffusion coefficient of $E$ has only two discrete states \red{\footnote{\red{
This discretization does not alter the system’s qualitative behavior. In fact, we verified that the demon effect remains qualitatively unchanged even when the four-state model is extended to include multiple motility states, allowing up to a tenfold increase in diffusivity.}}}. 
The $2 \times 2$ states of $E$ are denoted by subscripts $S$ and $P$ for the reactants, and superscripts $h$ and $l$ for motility. 
We consider stochastic dynamics within this subsystem, where $E$ can change its state by catalyzing a reaction, changing its motility, or leaving from the reactant. 
The state transitions are illustrated in \figref{fig:demonPathways}. 
When $E$ overlaps with $S$ or $P$, reaction occurs with a rate $a$ or $\kappa a$, respectively, and then the reactant switches to $P$ or $S$. 
After catalyzing $S$, $E$ changes into high-motility state, until it returns to the low-motility state with a rate $\gamma$. 
$E$ leaves from the reactant with a rate depending on its motility; for low-motility (high-motility) state, $\tilde{D}^l$ ($\tilde{D}^h$), respectively. 
(For more details, see supplementary materials including the calculation of $\tilde{D}^h$ and $\tilde{D}^l$.) 

This simplification highlights the presence of a typical pathway (red arrows in \figref{fig:demonPathways}A): 
When $E$ catalyzes a reaction from $S$ to $P$, its motility becomes high, and $E$ is more likely to leave from $P$. 
Then, the probability that $E$ catalyzes the reverse reaction decreases. 
In contrast, when $E$ catalyzes a reaction from $P$ to $S$, the probability to catalyze $S$ again remains unchanged. 
By this bias in the probability of consecutive reactions, the steady state deviates from the chemical equilibrium. 

Accordingly, $E$ plays a role of Maxwell's demon, which observes the state of the reactant, stores the information of that as the enhanced diffusion state, and changes the probability of following reactions. 
Thus, $E$ acts in the same way as Maxwell's demon to reduce the entropy of the system, which is achieved by the demon's motion. 
With EED, even if the catalytic reactions of enzymes do not change the chemical equilibrium, by using the information, enzymes can extract work from the system. 

If all reactions followed the optimal pathway highlighted by the red arrows in \figref{fig:demonPathways}A, one bit of information would be generated by the pathway, i.e., 
$E$ ``judges'' if the reactant is $S$ or $P$ and when it is $S$, $E$ turns it to $P$ and leaves from it and cease the reaction, and when it is $P$, $E$ turns it to $S$ but does not leave and continue the reaction to turn it to $P$. 
\red{This is shown in \figref{fig:informationLimits}A, where the deviation from equilibrium converges to 1 bit of information as consumed energy $\Delta K$ and reaction rate $a$ approach infinity.
The convergence to 1 bit of information implies that the demon can perfectly observe the state of the reactant and use that information to selectively leave $P$.}
\red{However, this ideal behavior is inhibited or disturbed by the following three factors:}
\begin{enumerate}
  \setlength{\leftskip}{5mm}
  \item $E$ leaves a reactant \red{regardless of its motility state; it does not utilize} the information \red{or change its behavior when the black dotted} arrows in \figref{fig:demonPathways} \red{are comparable in frequency to the red pathway}.
  This is dominant for $\tilde{D}^l \gg a$. 
  \item $E$ loses its enhanced motility before it leaves from the reactant; $E$ loses the memory fast before work extraction, (depicted by \red{blue dotted} arrows in \figref{fig:demonPathways}). 
  This is dominant where $\tilde{D}^h \frac{\kappa}{1+\kappa} a \gg \tilde{D}^l \gamma$, as the ratio of probability in high-motility state to low-motility state is roughly approximated to $\frac{\kappa}{1+\kappa} a / \gamma$. 
  This corresponds to the case where the leaving of $E$ at the low-motility state from a reactant dominates that of $E$ at the high-motility state. 
  \item The reverse reaction occurs before $E$ leaves from $P$; $E$ observes a reactant as demon but cannot convert it to work, and finally information is lost (depicted by \red{green dotted} arrows in \figref{fig:demonPathways}). 
  This is dominant for $\kappa a \gg \tilde D^h, \gamma$. 
\end{enumerate}
The system exhibits the deviation from equilibrium when the above three conditions are not satisfied. 
Indeed, by drawing phase diagram of the deviation from equilibrium by calculating the steady state analytically with each parameter set, the deviation from equilibrium is observed only within the region bounded by the above boundaries (\figref{fig:analysis}). 
\red{Even under this region, there remain pathways that that interfere with the demon’s transfer of input energy into information. 
Assuming that the energy input per $S \rightarrow P$ reaction is $\Delta K$ and information gain is difined by Kullback-Leibler divergence between the reactant state distribution with and without EED, the demonn's transfer efficiency is computed, as shown in \figref{fig:informationLimits}B.}

Note that some discrepancies remain between the phase diagram obtained from the particle simulation (\figref{fig:simulationResults}C) and that obtained from the analytical calculation of the four-state model (\figref{fig:analysis}). 
One discrepancy appears when $A$ is large and $V$ is small, in which the deviation from equilibrium is observed in the four-state model but not in the particle simulation. 
This is because, in the four-state model, we ignore the multiplicity of reactants; we neglect the possibility that $E$ in high-motility state leaving $P$ overlaps with other reactants before returning to the low-motility state. 
In contrast, in the particle simulation, this effect is not ignored. 
Once the $E$ with enhanced motility leaves the reactant, it can traverse a distance of $\sqrt{ \frac{\Delta K A + 1}{V}}$ before returning to the low-motility state. 
If this distance is larger than the average distance between reactants $({\rho_S + \rho_P})^{-\frac{1}{2}}$, i.e., $8 (\Delta K A + 1) (\rho_S + \rho_P) r^2 > V $, high-motility $E$ overlaps with the other random reactant, then the information is lost. 
It corresponds to the area below the gray line in \figref{fig:simulationResults}C. 
Indeed, when the density of particles is decreased, this effect of multiplicity is weakened. 
Then, the region with the deviation from equilibrium is expanded, as is numerically confirmed (see \supfigref{fig:densityDependence}).

The other discrepancy appears where $A$ and $V$ are large; the deviation from equilibrium is smaller in the four-state model (\figref{fig:analysis}) than that in the particle simulation (\figref{fig:simulationResults}C).
This is due to the simplification of the motility of $E$ to two states in the four-state model. 
In the particle simulation, the motility of $E$ continuously changes, and can be accumulated up to larger value than $\Delta K$. 
When $A$ and $V$ are large, $E$ hardly leaves from the reactant, and reactions can sequentially occur.
Then the motility is continuously enhanced before it decays. 
This effect is not considered in the four-state model, and the deviation from equilibrium is underestimated. 

Note that the deviation from equilibrium also can be observed for $\kappa \neq 1$ (\figref{fig:kappaVSratio}). 
In the equilibrium, the ratio of the number of $S$ and $P$ particles is precisely equal to $1/\kappa$, i.e., $\rho_P / \rho_S = 1/\kappa$.
However, the ratio of the concentration of $S$ and $P$ particles is deviated from the equilibrium with EED for any $\kappa$, $\rho_P$ always increases and $\rho_S$ decreases. 
The magnitude of the deviation is not additive but multiplicative, i.e., the ratio of $\rho_P / \rho_S$ to $1/\kappa$ takes a similar value for all $\kappa$.

In this Letter, we have shown that EED leads to a deviation in steady state concentrations away from chemical equilibrium. 
This phenomenon arises because the enzyme uses information about the type of reactant, obtained via EED, to suppress subsequent reactions with the same reactant. 
This behavior can be interpreted as a manifestation of Maxwell's demon. 

Although we assumed the increase in diffusion coefficient as a form of EED in the particle-reaction model, our findings are not limited to cases with changes in diffusion coefficient. 
Recent studies have also shown that EED involves anomalous diffusion or directed motion like chemitaxis \cite{Jee2018-ys, Lin2022-ct, Vishen2024-qu}. 
Even in those cases, our mechanism still works as long as the probability of the enzyme to leave from the product is enhanced \red{\footnote{\red{The demon effect presented here does not depend on how EED is implemented. 
For instance, the origin of $\Delta K$ in our model can be either an increase in hydrodynamic motility due to stochastic swimming or an effective diffusivity boost caused by transient temperature rises, as discussed in the models of \cite{Golestanian2015-lt}.
In the former case, $\Delta K$ in our model can be interpreted as the fractional increase in hydrodynamic motility. 
In the latter case, heating, $\Delta K$ arises from transient temperature increases following exothermic catalysis. 
This can be formalized by defining an energy release $\Delta E = 1 / \gamma \tau_d$ per reaction, with a dissipation timescale $1/\gamma$ and the timescale of enzyme’s local temperature relaxation $\tau_d$.}}}. 

The parameter range where the deviation from equilibrium is observed aligns with experimentally measured parameters in enzymes showing EED; as for $A$, it should be between $10^{-1}$ and $10^1$. 
It has been reported that this condition can be satisfied by enzymes showing EED. 
For instance, in urease\footnote{As far as we know, quantitative measurement of  $\gamma$ is available only for urease, so that we examined the condition for it.}, the turnover rate is on the order of $10^4$ to $10^5$ $\text{s}^{-1}$, meaning that one catalytic cycle takes approximately 10 to 100 $\mu \text{s}$ \cite{Krajewska2009-wz}, where EED is estimated to last about 6 $\mu s$ \cite{Jee2018-hp, Jee2018-ys}. 
Therefore, $A$ value estimated from this experimental data is on the order of $10^0$ to $10^1$, which is consistent with our theoretical prediction. 

On the other hand, the parameter $V$ is given by $V = 4 \gamma \nu r^2$, with the parameter of the viscosity $\nu$, the radius of the enzyme $r$, and the decay rate of the motility $\gamma$. 
This parameter represents the relative timescale of the natural diffusion of the enzyme within the particle radius compared to the timescale of the EED. 
In the case of urease, the radius of the enzyme is on the order of 10 nm \cite{Mobley1995-jr}, and the diffusion coefficient of urease is estimated as roughly $10^{-7} \text{cm}^2/\text{s}$ \cite{Muddana2010-yi}. 
Therefore, the experimental value of $V$ for urease is calculated $\approx 1$, which is close to the region of the deviation from equilibrium in the phase diagram \footnote{While this phenomenon is not observed with $V = 1$ in the shown data with the particle simulation (\figref{fig:simulationResults}C), we enphasize that this is due to the density effect of particles, and the deviation from equilibrium arises with $V = 1$ when the density of particles is decreased, as calculated in the four-state model and numerically confirmed in \supfigref{fig:densityDependence}.}.  
Hence, \red{the} enhanced product concentration \red{caused} by EED is expected to work from the experimentally observed data\red{, and also can be examined by in-vitro experiments. 
For instance, the viscosity dependence of demon effect in \figref{fig:simulationResults} and \figref{fig:analysis} can be examined by changing the concentration of the solution.}

\red{In real biological systems, many additional factors, such as the inflow and outflow of particles and molecular crowding, are present. 
We have simulated models incorporating these factors and confirmed that the mechanism of EED acting as a Maxwell’s demon remains valid, as shown in \supfigref{fig:otherExamples}. 
This is reasonable, as the effect occurs shortly after the reaction and acts locally between a single reactant and an enzyme.}

This study has demonstrated that EED, previously explored in the context of chemotaxis and compartmentalization \cite{Dey2014-mi, Wu2015-fs, Weistuch2018-hk, Jee2018-hp, Jee2018-ys, Mohajerani2018-xt, Giunta2020-bm, Lin2022-ct}, can also induce nonequilibrium \red{steady} states \red{with concentration deviation from the equilibrium, even in the absence of external substrate and product flows.
This occurs because the enzyme, when undergoing EED, effectively acts as a Maxwell’s demon}. 
Unlike traditional discussions on proteins as Maxwell's demon, \red{for} molecular machines \cite{Toyabe2010-hc, Mizraji2021-vv}, \red{the} enzymes preserve information on the previous reaction in the form of enhanced diffusion and effectively transfer it into macroscopic nonequilibrium behavior. 
This novel perspective not only enhances our fundamental understanding of enzyme function but also provides a potential bridge between microscopic enzymatic activity and macroscopic biological nonequilibrium systems, such as those found in living organisms.

\begin{acknowledgments}
\paragraph{Acknowledgments.} 
This work is supported by RIKEN Junior Research Associate Program (to S.I.), Japan Society for the Promotion of Science (JSPS) KAKENHI Grant No. JP22K21344 (to S.I., T.S.H.), JP21K15048 (to T.S.H), and Novo Nordisk Foundation Grant No. NNF21OC0065542 (to K.K.).
\end{acknowledgments}

\bibliography{ref}

\clearpage
\widetext
\begin{center}
\textbf{\large Supplemental Materials}
\end{center}
\setcounter{equation}{0}
\setcounter{figure}{0}
\setcounter{table}{0}
\setcounter{page}{1}
\makeatletter
\renewcommand{\theequation}{S\arabic{equation}}
\renewcommand{\thefigure}{S\arabic{figure}}
\renewcommand{\bibnumfmt}[1]{[S#1]}
\renewcommand{\citenumfont}[1]{S#1}
\newcommand{\Var}{\mathrm{Var}}
\newcommand{\Cov}{\mathrm{Cov}}

\section{the four-state model} \label{supsec:derivationAnalyticalModel}

We derive a four-state model that describes the system of $S$, $E$, and $P$ particles. 
We assume the diffusion coefficient of $E$ is discretized into two states, high motility and low motility state with diffusion coefficients given by $\Delta K + 1/\nu$ and $1/\nu$, respectively. 
\red{This discretization should not change the behavior of the system, because the continuous diffusion coefficient and the discretized diffusion coefficient are equivalent in the sense that the distribution of distances that the enzyme $E$ can traverse after the reaction is the same in both cases.
With this formulation}, we represent the state of $S$ and $P$ particles by the number of overlapping $E$ particles with high and low motility.

For the simplest case, we consider the system focusing on the relationship between two particles, $S$ and $E$, or $P$ and $E$. 
When $E$ overlaps with the other particle, we can describe the system as any of 4 states, $E$ at high motility state overlapping with $S$, $E$ at low motility state overlapping with $S$, $E$ at high motility state overlapping with $P$, and $E$ at low motility state overlapping with $P$. 
We denote the probability of $E$ taking these states as $P_S^h, P_S^l, P_P^h$, and $P_P^l$, respectively. 
The schematic diagram of transition is illustrated as arrows in the right side of \figref{fig:demonPathways}.
The time evolution of these probabilities can be described by the master equations as follows.
\begin{align} \label{eqn:masterEquation}
  \dv{t} P_S^h &= - a P_S^h + \kappa a P_P^h - \gamma P_S^h - \tilde{D}^h P_S^h \nonumber \\
  \dv{t} P_S^l &= - a P_S^l + \kappa a P_P^l + \gamma P_S^h - \tilde{D}^l P_S^l + J_S \nonumber \\
  \dv{t} P_P^h &= - \kappa a P_P^h - \gamma P_P^h - \tilde{D}^h P_P^h \nonumber \\
  \dv{t} P_P^l &= a P_S^h  + a P_S^l - \kappa a P_P^l + \gamma P_P^h - \tilde{D}^l P_P^l + J_P
\end{align}
Here, $J_S$ and $J_P$ are the flux of $E$ particles in low motility state entering the overlapping state with $S$ and $P$ particles, respectively. 
They are determined to be consistent to the conservation of $E$ particles, $J_S = \tilde{D}^l P_S^l + \tilde{D}^h P_S^h$ and $J_P = \tilde{D}^l P_P^l + \tilde{D}^h P_P^h$. 
In other words, there are only one $S$ or $P$ and one $E$ particle in the considering system, therefore, when $E$ does not overlap with $S$ or $P$, 
The terms $\tilde{D}^h$ and $\tilde{D}^l$ are the rates at which $E$ particles in high and low motility states exit the overlapping state, respectively, and are estimated in next section. 

Obtaining the steady solution of these equations, we get the parameter-dependence of the deviation from the equilibrium as shown in \figref{fig:analysis}.

\subsection{Approximation of the escape rate of $E$ particles}\label{supsec:escapeRate}

Here, we estimate the escape rate of $E$ particles from the overlapping state. 
Because we consider only two particles in the system, the system can be described by the relative position between the two particles and the diffusion is described by the Brownian motion of it. 
In that sense, the diffusion rate of the distance between $E$ and $S$ or $P$ particles is sum of the diffusion rates of $E$ and $S$ or $P$ particles, when $E$ is in high motility state, $D^h = (\Delta K + 2) / \nu$, and when $E$ is in low motility state, $D^l = 2 / \nu$. 

Because particle motion follows Brownian motion and the typical displacement is proportional to the square root of time, it is difficult to precisely estimate the escape rate as a linear time constant. 
Here, we estimate the escape rate in a straightforward and simple manner by taking the inverse of the typical time it takes for two particles to transition from a completely overlapping state to a non-overlapping state.

With low-motility $E$ particles, the diffusion rate is $D^l = 2 / \nu$ and the typical time to move the distance of the sum of the radii of the two particles, $2r$, is $(2r)^2 / 2D^l = r^2 \nu$.
Therefore, the escape rate of low-motility $E$ particles is estimated as $\tilde{D}^l = 1 / r^2 \nu $.
For high-motility $E$ particles, as the diffusion rate is $D^h =(\Delta K + 2) / \nu$, and then, the escape rate of high-motility $E$ particles is estimated as $\tilde{D}^h = (\Delta K + 2) / 2r^2 \nu$.

\section{density dependence of the deviation from equilibrium} \label{supsec:reactantsDensityDependence}

As mentioned in the main text, the four-state model does not consider the multiplicity of particles, therefore, the model does not hold when the density of reactants is too high. 
We estimate the density dependence of the deviation from equilibrium by comparing the typical length for $E$ in high motility state to move before it returns to the low-motility state and the average distance between reactants. 
When the typical length is larger than the average distance, enzymes in high motility state can enter the overlapping state with the other random reactant before returning to the low-motility state.
In this case, the correlation between the state of the reactant and the motility of $E$ is lost, and the deviation from equilibrium is weakened. 
$E$ continuously reacting with the same reactant has the diffusion coefficient of $\Delta K A + 1/\nu$ on average, because the motility of $E$ is enhanced by $\Delta K$ with rate $a$ and decays with rate $\gamma$. 
Once it leaves the reactant, it typically moves the distance $\sqrt{\frac{\Delta K a/\gamma + 1}{\gamma \nu}}$ before its motility returns to the low-motility state with the decay rate $\gamma$.
If this length is larger than the average distance between reactants, $(\rho_S + \rho_P)^{-\frac{1}{2}}$, the information is disturbed by high-motility $E$ overlapping with another reactant.
Therefore, the boundary where this analytical model is invalid is estimated as $V = (\rho_S + \rho_P) (\Delta K A + 1)$.

\begin{figure}[tb]
  \centering
  \includegraphics[width=\columnwidth]{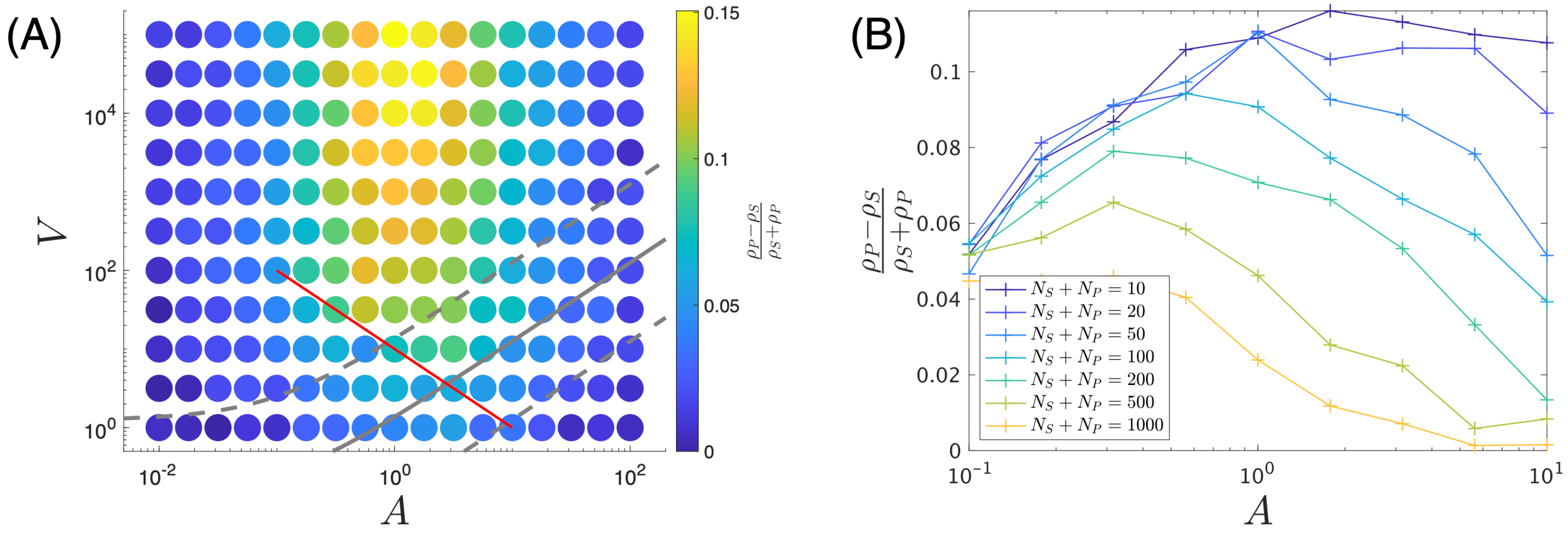}
  \caption{(A) Lines of the limitation of the four-state model with different number of particles and the range of parameters are plotted on the phase diagram of the deviation from equilibrium used in the main text. 
  The simulation conducted in main text has $1000$ particles of $S$ and $P$, and the limitation line of the four-state model is shown as the grey solid line. 
  As the number of particles decreases, the line declines to bottom right, and by changing the number of particles from $1000$ to $10$, the line moves from dashed line on top to bottom right. 
  The red line represents the line along which the data shown in (B) is calculated.
  (B) The dependence of the deviation from equilibrium on the number of reactants particles. 
  When the number of particles is high, the deviation from equilibrium is still observed, but decreases at high large $A$. 
  As the number of particles decreases, the deviation from equilibrium becomes observed in wider range of parameters as the line declines to right. 
  }
  \label{fig:densityDependence}
\end{figure}

With this estimation, we expected that the deviation from equilibrium is observed in wider range of parameters as the number of particles decreases, especially in the region where $A$ is large and $V$ is small.
To verify this expectation, we conducted the simulation with different number of particles and plotted the change of the deviation from equilibrium as functions of $A$ and $V$ (\figref{fig:densityDependence}). 
As the number of particles decreases, the area where the deviation from equilibrium is observed extends to the region with larger $A$ and smaller $V$, as expected from the above estimation of the boundary of the multiplicity of particles.

\begin{figure}[tb]
  \centering
  \includegraphics[width=0.6\columnwidth]{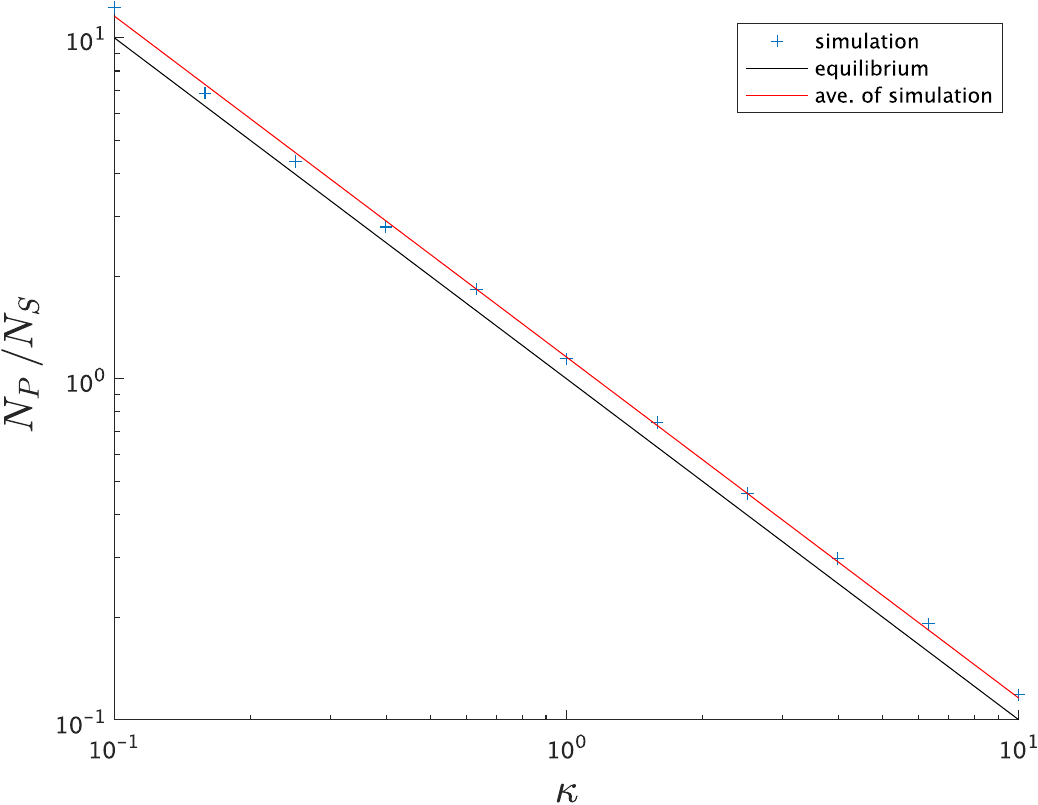}
  \caption{The dependency of the ratio of the number of $S$ and $P$ particles on the reversibility of the reaction. 
  The black line represents the equilibrium ratio of the number of $S$ and $P$ particles, $1/\kappa$. 
  The simulation data, performed with EED and represented by cross marks, show the deviation from equilibrium. 
  All simulations are performed with $a = 1, \gamma = 1, \nu = 10^5$ and $\Delta K = 10$. 
  The red line is a fit to the simulation data and average of the change of the ratio is $15 \%$.}
  \label{fig:kappaVSratio}
\end{figure}

\section{the particle-reaction simulation with more realistic features} \label{supsec:real}
\red{
In the main text, we showed the results of particle-reaction simulations using the simplest model, without repulsive interactions due to collisions between particles in an isolated system.
In this section, we present simulation results including collisions between particles and substrate uptake flow from the boundary, to better model cellular environments.
These features do not change the qualitative behavior of the system.}

\red{First, we consider substrate inflow and reactant outflow at the system boundaries.
When reactants reach the boundary, they are removed from the system, and the same number of reactants are added at random positions on the boundary.
Enzymes are not removed, so a closed boundary condition is applied to them.}

\begin{figure}[tb]
  \centering
  \includegraphics[width=\columnwidth]{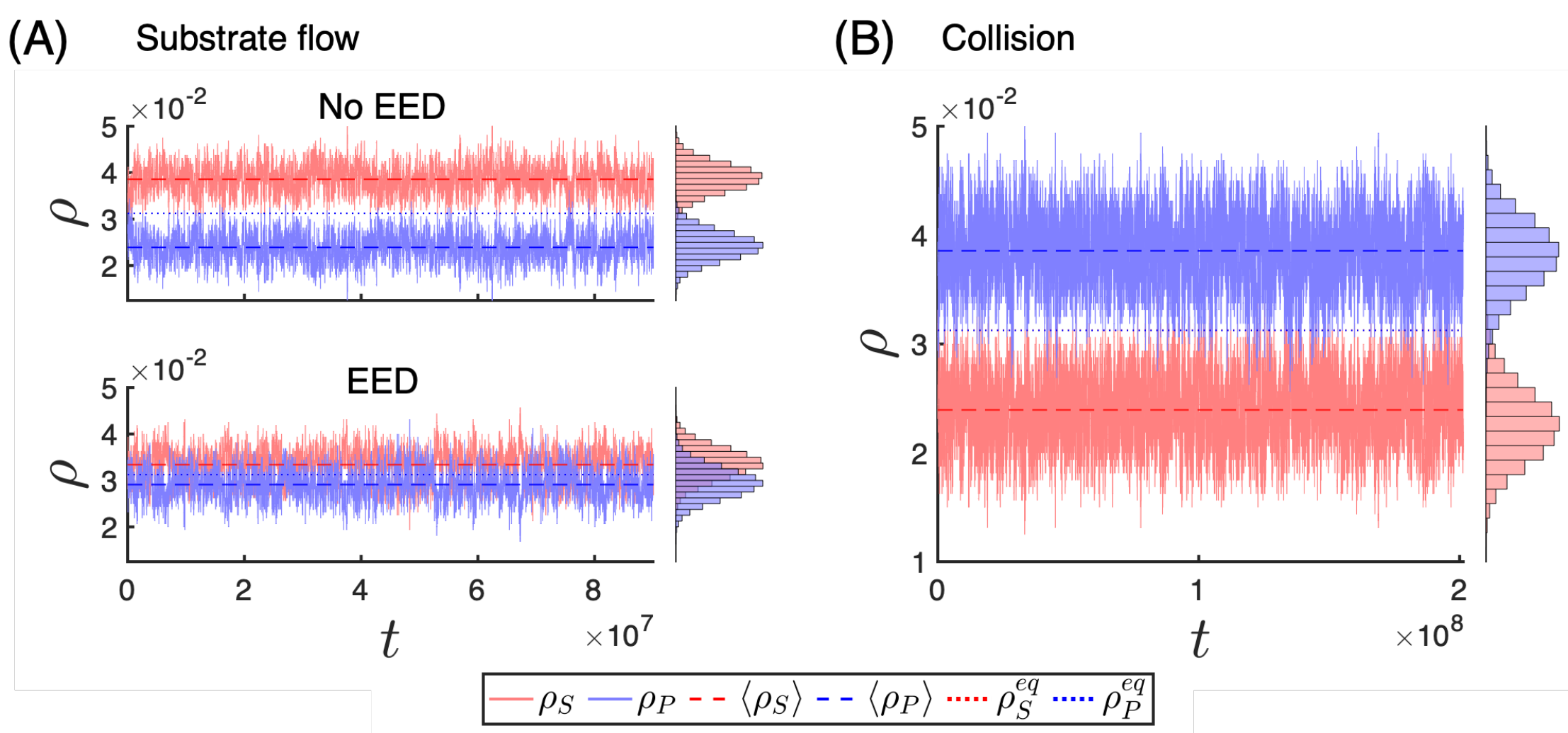}
  \caption{\red{(A) The time series and distributions of the concentrations of S and P in the particle-reaction simulation with substrate inflow from the boundary. 
  The upper panel shows the case without EED, while the lower panel shows the case with EED. 
  Solid lines represent the time series of $\rho_S$ and $\rho_P$ in the simulation. 
  Dotted lines indicate the values of $\rho_S$ and $\rho_P$ calculated at chemical equilibrium. 
  Dashed lines show the time-averaged concentrations.
  (B) The time series and the distribution of the concentration of $S$ and $P$ in the particle-reaction simulation with the Lennard-Jones potential with a cutoff.}
  }
  \label{fig:otherExamples}
\end{figure}

\red{When EED is not implemented, the system converges to a steady state where the concentration of $S$ is higher than that of $P$. This is due to the supply of $S$ particles from the boundary, as shown by the red and blue lines for $S$ and $P$, respectively, in the upper panel of \figref{fig:otherExamples}A. 
When EED is implemented, the system reaches a steady state in which the concentration of $P$ is higher than in the no-EED case, as shown in the lower panel \figref{fig:otherExamples}A. 
This demonstrates that EED can increase the concentration of $P$, even in systems with substrate inflow at the boundary.}

\red{
Next, we consider the collision between particles with the Lennard-Jones potential with a cutoff. 
The potential is given by
\begin{equation}
  V(r) = \begin{cases}
    4 \epsilon \left[ \left( \frac{\sigma}{r} \right)^{12} - \left( \frac{\sigma}{r} \right)^{6} \right] & (r < r_c) \\
    0 & (r \geq r_c)
  \end{cases}
\end{equation}
where $r_c$ is the cutoff distance, as the sum of the radii of two particles, $r_c = 1$. 
The parameters $\epsilon$ and $\sigma$ are set to $10^{-2}$ and $2^{1/6}$, respectively, so that the potential is only repulsive.
}
\red{
With this repulsive potential, the concentration of $P$ is larger than that of $S$ for the steady state, as shown in \figref{fig:otherExamples}B.
}

\section{Enhanced enzyme diffusion as information processing}

\red{As mentioned in the main text, enhanced enzyme diffusion (EED) can be interpreted as a form of information processing. In this section, we discuss the information gained and the efficiency of the demon.}

\begin{figure}[tb]
  \centering
  \includegraphics[width=\columnwidth]{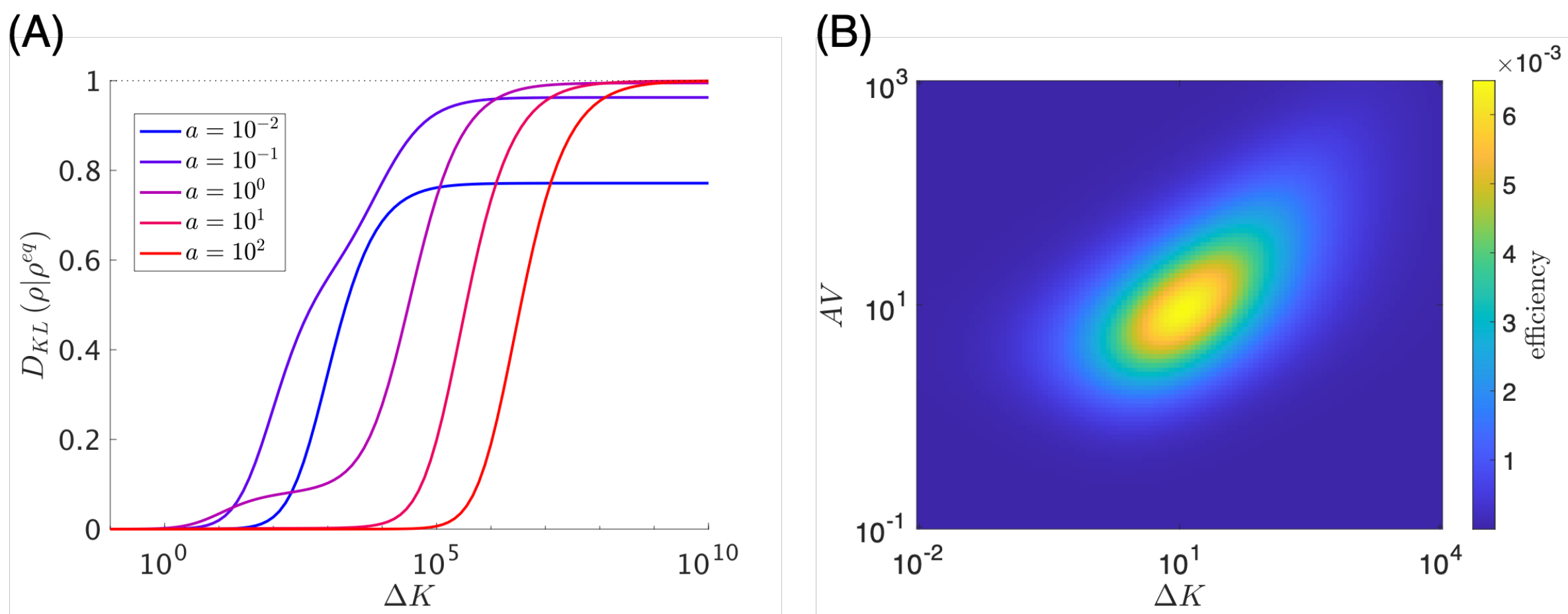}
  \caption{\red{(A) The Kullback-Leibler divergence $D_{KL}$ between the steady-state reactant distribution with EED and the chemical equilibrium distribution is plotted as a function of $\Delta K$ and $a$. 
  (B) The efficiency of the information processing, defined as the ratio of the Kullback-Leibler divergence to the consumed energy.}}
  \label{fig:informationLimits}
\end{figure}

\red{The enzyme, acting as a demon, observes the state of the reactant and alters its subsequent behavior, that is, whether to leave the reactant or not. 
In an extreme case, given infinite energy $\Delta K$ and an infinite reaction rate $a$, the demon works perfectly. 
In this case, the deviation from equilibrium converges to 1 bit, measured by the Kullback-Leibler divergence, which is the maximum information obtainable from the state of the reactant, as shown in \figref{fig:informationLimits}A. 
In this limit, however, the energy consumed by the demon also becomes infinite, meaning that infinite energy is required to gain 1 bit of information.}

\red{We then calculate the efficiency of the demon's information processing. To do this, we consider a function that describes the change in the reactant distribution caused by a single event: 
the enzyme $E$ overlaps with a reactant (either $S$ or $P$), catalyzes some reactions, and then leaves the reactant. 
This function maps the reactant distribution before the event to the distribution after the event. 
The four-state model can be derived by repeated application of this function, where the reactant distribution eventually converges to the steady state.}

\red{With this function, the cycle of the demon can be formulated as following:} 
\red{
\begin{enumerate}
  \setlength{\leftskip}{5mm}
  \item An enzyme $E$ in the low-motility state begins overlapping with a reactant, either $S$ or $P$, with a distribution corresponding to the equilibrium concentrations.
  \item Reactions occur and energy is consumed for EED upon the forward reaction.
  \item The enzyme leaves from the reactant with a biased distribution of the state of the reactant. 
  \item The distribution of reactants is returned to the chemical equilibrium to extract the work from the system.
\end{enumerate}
With a system temperature $T$, an energy of $\Delta K k_B T$ is consumed per reaction for EED. 
Hence, during the process, the total consumed energy is $\Delta K k_B T$ multiplied by the average number of the forward reactions.
When the distribution of the reactants is not at chemical equilibrium, work can be extracted by returning the distribution to chemical equilibrium. 
The amount of the extracted work is calculated as $k_B T$ multiplied by the Kullback-Leibler divergence the reactant distribution after the process and that at chemical equilibrium. 
The efficiency of the information processing is defined as the ratio of the extracted work to the consumed energy.}

\red{With this formulation, we can calculate the efficiency of the information processing and its parameter dependence by taking the limit of $\gamma \to 0$ (\figref{fig:informationLimits}B). 
The maximum efficiency is approximately 0.006, which is small but nonzero. 
This inefficiency arises because at least one of the two side routes, depicted by the black and green dotted arrows in \figref{fig:demonPathways}A cannot be avoided under finite energy input. 
To achieve perfect information processing, the demon must block the black route by suppressing $E$'s leaving from the reactant when it does not have enhanced motility.
This requires $AV=\infty$. 
However, with finite energy input, the leaving of $E$ from the reactant with high motility is also suppressed and the dominance of the green route increases, which results in the inefficiency of the information processing. 
Thus, the efficiency of the information processing is low even in the maximum case.}

\red{Moreover, as seen in \figref{fig:informationLimits}B, the efficiency of the information processing has a peak around $\Delta K \approx 10$ and $AV \approx 10$, which is indeed in the parameter range where enzymes can exhibit EED in experiments as discussed in the main text.}

\end{document}